# A large-scale dataset of sub-institution name disambiguation and hierarchical structures from OpenAlex

Zhentao Liang[a,b], Jin Mao[a,b,*], Gang Li[a,b]

[a] Center for Studies of Information Resources, Wuhan University, Bayi Rd 299, Wuhan 430072, China

[b] School of Information Management, Wuhan University, Bayi Rd 299, Wuhan 430072, China

* Corresponding author. Email: maojin@whu.edu.cn

**Abstract:**

Accurate attribution of scholarly work to specific sub-institutional units, such as schools or departments of a university, is crucial for granular research assessment and policymaking. While robust identifiers exist for top-level institutions, standardized data for sub-level units remains scarce due to the linguistic and structural variability of affiliation strings. In this study, we introduce OpenSubAffil, a large-scale dataset mapping raw affiliation strings from OpenAlex to disambiguated sub-institutional entities and their hierarchical structures. We developed a pipeline integrating named entity recognition (NER) with embedding-based clustering. Furthermore, we proposed a multi-signal scoring function that synthesizes lexical and co-occurrence evidence to reconstruct the sub-institutional hierarchy. OpenSubAffil comprises mappings for 40 million affiliation strings to 638,843 disambiguated sub-units across 18,635 top-level institutions, together with their hierarchical relationships. Validation against Wikidata benchmarks and manual investigation show that our method achieves promising performance. Overall, this dataset bridges the granularity gap between individual researchers and top-level institutions, enabling high-resolution analyses of scholarly output and communication at the sub-institutional level.

## Background & Summary

With an expanding body of scientific publications[1–3], accurately linking research output to the entities that produced it is essential for research evaluation and science policymaking[4–6]. The attribution of scholarly work to institutions, in particular, supports a wide range of applications, including university rankings[7–9], national R&D monitoring[10,11], funding allocation[12,13], and analysis of scholarly communication[14–16]. Achieving reliable institutional attribution at scale requires resolving the diverse and often inconsistent organizational references found in bibliographic metadata, a task known as institution name disambiguation[17,18]. Consequently, institution name disambiguation (IND) constitutes a foundational component of modern research information infrastructure.

In the past decade, significant progress has been made in disambiguating top-level institutions, such as universities and research institutes. Notably, the Research Organization Registry (ROR) provides more than 116,000 persistent identifiers for top-level research organizations, which serve as a standardized reference for resolving affiliation data[19]. Building on this foundation, open scientific knowledge graphs such as OpenAlex[20], Semantic Scholar[21], and OpenAIRE[22] have integrated IND into their data pipelines, linking raw affiliation strings to ROR identifiers. Moreover, ROR identifiers are also gaining adoption across the broader scholarly communication infrastructure. Crossref now supports ROR in its metadata schema[23], and ORCID has enabled researchers to associate their records with ROR-identified institutions[24]. Alongside these infrastructures, a growing body of dedicated disambiguation methods has emerged. For example, AffilGood[18] provides a pipeline that integrates span identification, named entity recognition, and entity linking to ROR. Entity-linking approaches based on edit distance and multi-stage retrieval-reranking architectures have also demonstrated promising performance[17,25,26]. As a result, top-level IND can now be performed with relatively high reliability at scale.

However, this progress has not extended to the sub-institutional level. Many analytical tasks require IND at a finer granularity, such as schools, departments, and research centers. For instance, departmental-level IND is essential for comparing research performance across organizational units within and between universities[27–29], and tracing collaboration and knowledge flows between these units[30–32]. Yet standardized data for these sub-level entities remain scarce. The ROR, by design, does not aim to comprehensively catalog university departments or subdivisions, as doing so would compromise its maintainability and scope (https://ror.org/about/faqs/#why-doesnt-ror-include-university-departments). While some national directories capture sub-institutional structures, such as GERiT for German research institutions[33] and the RNSR for French research units[34], these resources are country-specific and limited in scope. Unlike top-level IND, where ROR provides a universal target that enables entity-linking approaches, no comparable global reference exists for sub-institutional entities. Consequently, existing disambiguation systems resolve affiliation strings to top-level organizations while discarding the sub-institutional information[18,35]. This leaves a significant granularity gap between individual researchers, who are increasingly well-identified through persistent identifiers such as ORCID, and the top-level institution to which their work is attributed[36]. Furthermore, sub-institutional units are not isolated entities but organized hierarchically, with departments nested within schools, schools within faculties, and so on. No large-scale dataset currently provides both disambiguated sub-institutional entities and their hierarchical relationships.

Several challenges make the sub-level IND particularly difficult. First, sub-institutional names exhibit significant linguistic variability. A single department may appear under different names due to abbreviation, translation, or reordering across publications[10,37]. Second, affiliation strings rarely encode a complete organizational hierarchy. Intermediate levels, such as faculties or schools, are frequently omitted, making it difficult to reconstruct the hierarchical relationships between sub-institutional units[35]. Finally, the absence of large-scale annotated datasets or evaluation benchmarks for sub-level IND has constrained both methodological development and reproducible comparison across different solutions[10,35].

To bridge this gap, we introduce OpenSubAffil, a large-scale dataset that maps raw affiliation strings from OpenAlex to disambiguated sub-institutional entities and their hierarchical relationships. OpenSubAffil comprises mappings for 40 million affiliation strings to 638,843 disambiguated sub-units across 18,635 top-level institutions, together with 638,843 hierarchical parent-child relationships. The dataset is derived from English-language affiliation records in the educational sector of OpenAlex, using a pipeline that integrates named entity recognition, embedding-based clustering, and a multi-signal scoring function to reconstruct the hierarchy. To evaluate the quality of the dataset, we propose using Wikidata-derived benchmarks and complementing them with manual investigation. Validation results show that our sub-level IND achieves a macro precision of 0.884 and a macro recall of 0.941, while the reconstructed hierarchies achieve a macro precision of 0.940 and a macro recall of 0.793, respectively. The complete dataset and code are publicly available. Overall, OpenSubAffil bridges the granularity gap between individual researchers and top-level institutions, providing a foundational resource for high-resolution analyses of scholarly output and organizational dynamics at the sub-institutional level.

# Methods

## Overview of the dataset construction

The construction of OpenSubAffil involves four main procedures (Figure 1). First, we extracted raw affiliation strings and their associated institutional linkages from OpenAlex, and applied preprocessing and filtering to ensure input quality. Second, a two-stage named entity recognition framework was employed to identify sub-institutional mentions within each affiliation string. The recognized entities were then assigned to their corresponding top-level institutions, which have already been disambiguated and linked to ROR identifiers by OpenAlex. Third, we disambiguated the extracted sub-institutional names through embedding-based agglomerative clustering, grouping variant surface forms of the same unit into canonical entities. Finally, we reconstructed the hierarchical parent-child relationships among the disambiguated entities using a multi-signal scoring function. This pipeline produces two major outputs: a mapping table linking 40 million raw affiliation strings to 638,843 disambiguated sub-institutional entities across 18,635 institutions, and a hierarchy table containing 638,843 parent-child relationships.

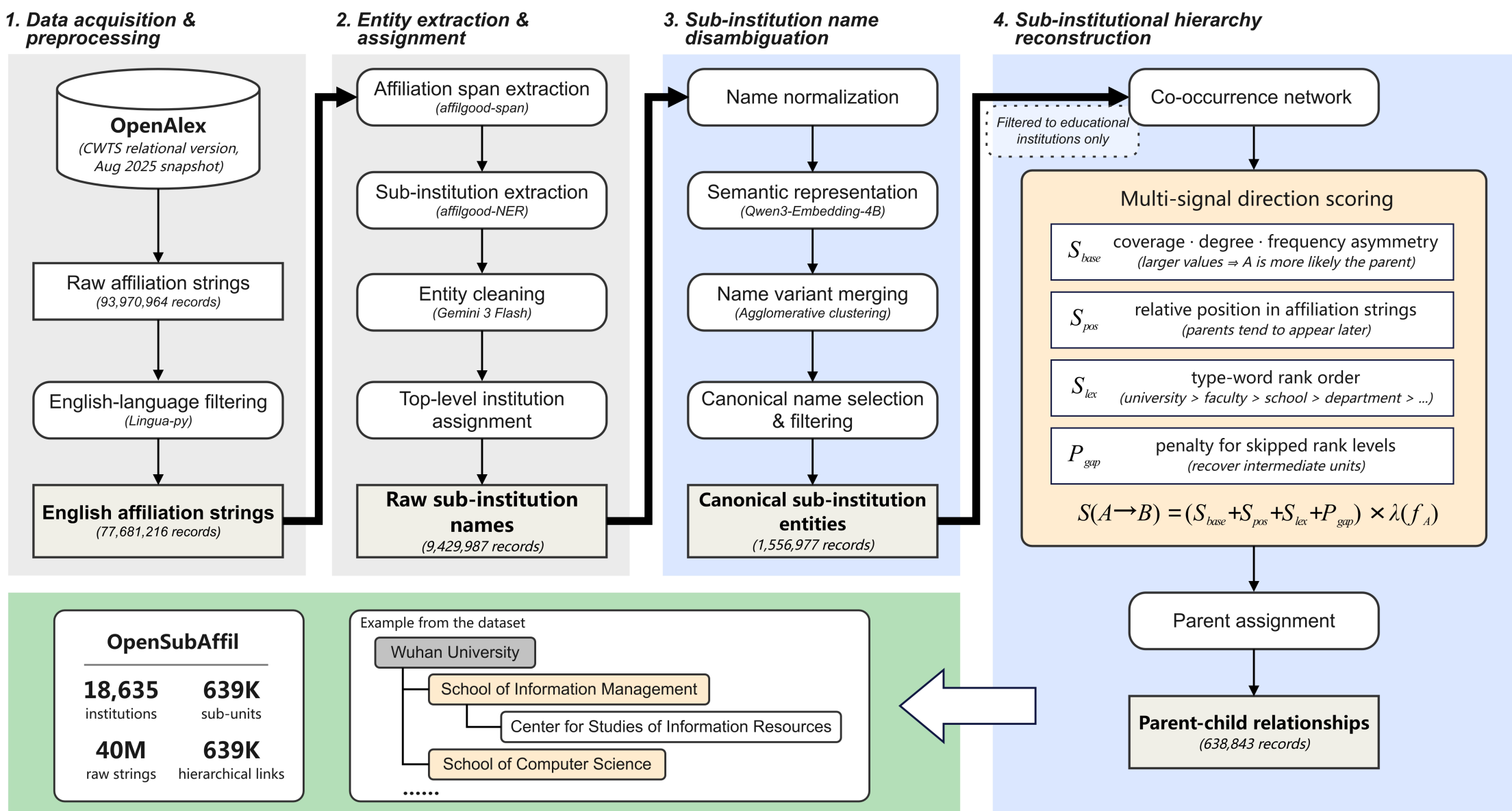

**Fig. 1** Overview of the dataset construction pipeline.

## Data acquisition and preprocessing

We sourced our data from the publicly accessible August 2025 snapshot of the CWTS version of the OpenAlex database, a relational database maintained by the Centre for Science and Technology Studies at Leiden University and hosted on Google BigQuery under the dataset identifier **cwts-leiden.openalex_2025aug**

(https://console.cloud.google.com/bigquery?ws=!1m5!1m4!3m2!1scwts-leiden!2sopenalex_2025aug!23sRESOURCE_LIST). We chose this version over the official OpenAlex JSON data dump for two reasons. First, OpenAlex updates its data on a weekly basis. While this is valuable for obtaining the latest publications, it makes reproducing analyses more difficult. The CWTS version addresses this by providing stable, versioned annual snapshots in a carefully designed relational format that facilitates large-scale analysis. In addition, CWTS maintains supplementary identifiers not present in the original OpenAlex data, including persistent identifiers for raw affiliation strings. This is essential to guarantee stable cross-version linkage and reproducible downstream analyses. The reliability of this database has also been validated through its use as the underlying data source for the renowned Leiden Ranking Open Edition (https://open.leidenranking.com/resources).

The August 2025 snapshot comprised 270 million scholarly works and 142 million raw affiliation strings. For each raw affiliation string, we queried the **work_affiliation**, **work_affiliation_institution**, and **raw_affiliation_string** tables to retrieve its associated top-level institutions as resolved by OpenAlex, along with its frequency of occurrence. We restricted our input to affiliation strings for which OpenAlex had already assigned at least one institution identifier, as top-level institution disambiguation in OpenAlex has reached a high accuracy[20] and is not the focus of our study. This yielded a total of 93,970, 964 raw affiliation string records. We then performed language detection using the Lingua-py library and retained only English-language records, as the downstream NER model used in our pipeline was trained predominantly on English text[18]. Note that this restriction has a limited impact on global coverage, as most affiliation strings are written in English regardless of the country of origin. The filtered dataset contains 77,681,216 records, accounting for 82.67% of the input.

### Entity extraction and top-level institution assignment

The goal of this stage is to extract raw sub-institutional name mentions from each affiliation string and assign them to the corresponding top-level institutions. These sub-institutional names serve as input for the subsequent disambiguation stage. We adopted the AffilGood framework[18] for this task, as it is one of the few open-source affiliation-parsing toolkits that explicitly distinguishes sub-institutional units (labeled SUB) from top-level organizations (labeled ORG). It has achieved an F1 score of 0.91 for SUB entity recognition.

The entity extraction follows a two-stage framework. Raw affiliation strings in OpenAlex are often noisy, containing extraneous elements such as author names, titles, email addresses, and numerical identifiers. A single string may also contain multiple institutional affiliations. Directly applying an NER model to such unprocessed strings degrades the recognition performance. Therefore, we first used the AffilGood span extraction model to identify and isolate clean, organization-relevant text segments from each raw string. In the second stage, we applied the AffilGood NER model to classify each extracted span into fine-grained entity types, retaining only those labeled as SUB. Not all affiliation strings yield SUB entities through this process. After extraction, 58,066,250 of the 77,681,216 input strings (74.75%) produced at least one recognized sub-institutional mention. The remaining strings either contained only top-level institution references or could not be parsed into meaningful sub-institutional entities.

The recognized entities then underwent further cleaning, including HTML entity decoding followed by lowercasing and punctuation removal. A particularly important step is abbreviation expansion (e.g., “dept comp sci” to “department of computer science”). As abbreviation forms are pervasive in affiliation strings, if left unresolved, they may cause the same unit to appear as distinct entities during disambiguation. To handle the large variety of abbreviation patterns, which is difficult to address exhaustively with handcrafted rules, we adopted an LLM-based approach. We first collected all SUB entities that contained common academic abbreviations (e.g., “dept”, “sch”, “inst”, “fac”) across the entire dataset, totaling 350 thousand distinct entities. Next, we used the Gemini 3 Flash model via its batch API to expand each abbreviated string into its full formal English form. The results were stored as a lookup dictionary and used during processing. The driver script, prompt template, and abbreviation-expansion lookup table used in this study are provided with the accompanying code and data release. We also filtered out SUB entities that exactly matched a known top-level institution name in ROR, as these were likely classification errors.

The final step of this stage is to assign each recognized SUB entity to its corresponding top-level institution. For affiliation strings associated with a single institution in OpenAlex, which account for 85.37% of all records, this assignment is straightforward. For strings associated with multiple institutions, we matched each SUB entity to its most likely parent by identifying the nearest co-occurring ORG entity in the string and comparing it against canonical institution names using fuzzy string matching. As a result, less than 4% of the records could not be reliably assigned and were excluded from subsequent processing. This stage produced 9,429,987 unique raw sub-institutional names across 66,332 top-level institutions, which served as input for the subsequent disambiguation stage.

## Sub-institutional name disambiguation

In this stage, we group variant surface forms of the same sub-institutional unit into clusters and assign each cluster a canonical name. Different from the top-level IND task, where a global registry like ROR provides a well-defined target for entity linking, no such reference exists for sub-institutional entities. We therefore frame this as an unsupervised clustering task, performed independently within each top-level institution. Specifically, we take the institutional assignments provided by OpenAlex as given and focus exclusively on resolving the finer-grained structure within each institution. In this way, OpenSubAffil complements rather than duplicates existing top-level IND efforts.

A key challenge is that raw sub-institutional names exhibit substantial surface variation even after the cleaning and expansion of abbreviations described above. In some cases, author names are erroneously attached to the beginning of sub-institutional name strings due to upstream parsing artifacts (e.g., John Smith School of Communication and Information). In addition, institutions in non-English-speaking countries often lack standardized English translations for their organizational levels, leading to the use of interchangeable terms such as school, college, and faculty to refer to the same unit.

To align these variant forms to their core names before clustering, we applied normalization rules within each top-level institution. To remove erroneous person-name prefixes, we split each name at the first occurrence of an organizational type word, e.g., department, school, and college, into a prefix and a suffix. A prefix is stripped only when frequency-based evidence within the same institution suggests that it is a rare variant rather than the predominant form of the name. Specifically, we require the suffix to appear independently with sufficient frequency, and the prefixed form to represent only a small fraction of the suffix’s total occurrences. This ensures that frequently used eponymous names, e.g., “Wee Kim Wee School of Communication and Information", are preserved. For type-word and modifier variation, we extracted the core name from each sub-institutional name by removing organizational type words and modifiers, so that "School of Computer Science", "College of Computer Science", and "Graduate School of Computer Science" yield the same core. Variants sharing the same core name within an institution were unified to the most frequent form. All normalization rules are governed by conservative frequency thresholds to prevent over-merging.

While the deterministic rules above have addressed systematic sources of variation, a long tail of residual differences remains, including misspellings, word reordering, and subtle paraphrasing. These variants are difficult to enumerate with rules, but are semantically similar and can be effectively captured by text embedding models. Therefore, we generated dense vector representations for all unique sub-institutional names after normalization using the Qwen3-Embedding-4B model, a SOTA embedding model on the Massive Text Embedding Benchmark[38]. Agglomerative clustering with average linkage was then performed independently within each institution, using a cosine distance threshold of 0.15. This threshold was determined through manual inspection of borderline cases near the decision boundary. No predetermined number of clusters was imposed. For each resulting cluster, we selected the most frequent sub-institutional name as the canonical name.

To further manage long-tail noise, we applied a coverage-based filtering strategy to keep only the most reliable canonical names. Specifically, canonical names within each institution were ranked by frequency in descending order, and only those cumulatively accounting for 95% of the raw affiliation strings were retained. After filtering, 1,556,977 canonical sub-institutional names across 66,332 top-level institutions were retained, covering 53,262,990 raw affiliation strings.

## Sub-institutional hierarchy reconstruction

Affiliation strings often list multiple sub-institutional entities alongside their parent institution in a single record, and the patterns of their co-occurrence carry informative signals about hierarchical structure. We exploit these signals to infer parent-child relationships among the disambiguated sub-institutional entities within each top-level institution. This results in a tree rooted at the institution itself.

For each top-level institution, we first constructed a co-occurrence network from the raw affiliation strings. Specifically, for each affiliation string mentioning multiple sub-institutional entities, we recorded all pairwise co-occurrences. Furthermore, we added the top-level institution itself as a root node connecting all sub-units into the co-occurrence network, so that the hierarchical relationships between the institution and its direct children can also be inferred.

Next, to infer the direction of hierarchical (parent-child) relationships, we designed a multi-signal scoring function that evaluates, for each co-occurring pair, the likelihood that one is the parent of the other. Given a pair $(A, B)$ where $A$ is the hypothesized parent, the direction score is computed as:

$$S(A \rightarrow B) = \left(S_{\text{base}} + S_{\text{pos}} + S_{\text{lex}} + P_{\text{gap}}\right) \times \lambda(f_A) \tag{1}$$

Each component captures a different aspect of the directional evidence. The base component $S_{\text{base}}$ serves as the foundation of the scoring function, which directly quantifies the asymmetry between two co-occurring entities using three statistical signals from the network:

$$S_{\text{base}} = w_c \cdot (c_{A|B} - c_{B|A}) + w_d \cdot \frac{d_A - d_B}{d_A + d_B} + w_f \cdot \frac{f_A - f_B}{f_A + f_B} \quad (2)$$

where $c_{A|B}$ is the proportion of $B$'s records in which $A$ also co-occurs and vice versa, $d$ denotes the node degree in the co-occurrence network, $f$ denotes the frequency of the entity, and $w$ denotes the weight applied to each term. The first term measures the difference in coverage. In practice, when authors list a department as their affiliation, they usually include its parent school or faculty, but the reverse is not necessarily true. Therefore, a large positive value of $(c_{A|B} - c_{B|A})$ suggests that $A$ is the parent of $B$. The second term captures the degree difference. Parent entities tend to co-occur with a wide range of other entities, reflecting their higher position in the organizational hierarchy. The third term captures the frequency difference, as parent entities generally appear in more records than their children. All three terms were normalized to the range [-1, 1] for comparability. We set $w_c = 0.4$, $w_d = 0.1$, and $w_f = 0.2$. These weights were determined heuristically during method development based on manual inspection of the reconstructed hierarchies, with greater emphasis placed on the difference in coverage as the most direct co-occurrence signal.

While $S_{\text{base}}$ effectively identifies the direction of a potential parent-child relationship, it alone tends to produce overly flat hierarchies, as top-level entities naturally dominate all three difference measures against most sub-units. This is a known limitation of co-occurrence-based methods[35]. To reconstruct the full depth of organizational hierarchies, especially for intermediate units such as faculties and schools, we introduce three complementary components.

The position component $S_{\text{pos}}$ provides ordering evidence independent of frequency. In affiliation strings, sub-units are conventionally listed from the specific to the general. Therefore, parent entities tend to appear later in the string. We capture this through two complementary measures:

$$S_{\text{pos}} = w_p \cdot (w_{po} \cdot P_{ord} + w_{pp} \cdot P_{avg}) \quad (3)$$

where $P_{ord}$ measures the proportion of co-occurring affiliation strings in which $B$ precedes $A$. While this is the direct measure of relative position, it may become less reliable when the co-occurrence count is small. To complement it, $P_{avg}$ calculates the difference between the mean normalized positions of $A$ and $B$ in affiliation strings. We set the outer position weight $w_p = 0.3$ and assigned equal internal weights to the two position measures. Together with the base-component weights, $w_c + w_d + w_f + w_p = 1$.

The lexical component $S_{\text{lex}}$ encodes prior knowledge about organizational type words, which follow a broadly consistent hierarchy across institutions worldwide. We define four rank levels: university (rank 1) > faculty (rank 2) > school/college (rank 3) > department/center/laboratory (rank 4). For each entity, we infer its rank from the highest-level type word present in its name. A bonus is awarded when the hypothesized parent holds a higher rank:

$$S_{\text{lex}} = \text{clip}\left(\frac{r_B - r_A}{|R|}, -L_{\text{cap}}, L_{\text{cap}}\right) \quad (4)$$

where $r$ denotes the inferred rank, $R = 4$ is the total number of rank levels, and $L_{\text{cap}} = 0.25$. The cap prevents the lexical prior from dominating the statistical and positional evidence. The lexical component and the rank-gap penalty described below are heuristic correction terms and are therefore not included in the normalized weighting of the primary evidence score.

However, $S_{\text{lex}}$ alone only encourages the correct direction but does not penalize relationships that skip intermediate levels. A direct link from university to department may reflect the common practice of omitting intermediate units in affiliation strings, rather than the actual absence of those units in the organizational structure[35]. When a subset of authors includes these intermediate units, the algorithm should be able to leverage this evidence to recover them. The rank gap penalty component $P_{gap}$ serves this purpose:

$$P_{\text{gap}} = -\alpha \cdot \max(r_B - r_A - 1, 0) \tag{5}$$

where $\alpha = 0.15$ controls the penalty strength per skipped level. For example, a direct link from university (rank 1) to department (rank 4) incurs a penalty of $-0.15 \times 2 = -0.30$, while a link from school (rank 3) to department (rank 4) incurs no penalty, making the latter substantially more attractive when both are available as candidates.

Finally, the parent frequency scaling factor dampens the overall score for high-frequency parent candidates:

$$\lambda(f_A) = \frac{1}{1 + log(1 + f_A)} \tag{6}$$

Without this correction, frequently occurring entities may dominate the scoring and attract disproportionately many children, regardless of the actual organizational structure.

After computing the directional scores, we applied a sequence of candidate-edge filters and selection rules. For each entity pair, let $n_{AB}$ denote the co-occurrence count and $f_A$ and $f_B$ denote the frequencies of the two entities. A pair was retained only when $n_{AB} \geq 2$ and its association score, defined as $n_{AB}/min(f_A, f_B)$, was at least 0.01. At the institution level, we then retained candidate directions with scores of at least $max(0, q_{0.5})$, where $q_{0.5}$ is the median of the candidate direction scores. For each child entity, at most three candidate parents were retained. The remaining edges were sorted by score and selected greedily, subject to the constraints that each child had at most one parent and that no directed cycle was created. Any canonical sub-institution that did not receive a selected parent was subsequently connected directly to the top-level institution during root completion. These rules served as conservative safeguards against low-support and weakly associated edges, while limiting competing parent candidates and enforcing a single-parent, acyclic hierarchy. These conservative rules reduce low-support and weakly associated candidate edges while enforcing a sparse, single-parent, acyclic hierarchy. Because entities without a surviving parent candidate are connected directly to the institutional root, the filtering procedure may also contribute to shallower reconstructed hierarchies, in addition to the omission of intermediate units from affiliation strings.

After these filtering and selection steps, we obtained 1,524,837 parent-child relationships across 66,332 top-level institutions, of which 314,812 are intermediate-level relationships.

The pipeline described above was applied to all institution types in OpenAlex. However, we decided to restrict the primary release of OpenSubAffil to educational institutions for two reasons. First, our pipeline is designed around the organizational structure of educational institutions, and its assumptions may not generalize to other types of institutions, such as companies, hospitals, or government agencies. Second, both the external benchmarks and manual verification used in our technical validation primarily cover educational institutions, making it difficult to assess data quality for other types with comparable rigor. After filtering, the final dataset comprises 18,635 educational institutions, 39,932,792 raw affiliation strings, and 638,843 parent-child relationships.

## Data Records

The OpenSubAffil dataset has been deposited on Zenodo and is publicly accessible[39]. The dataset adopts a normalized relational design comprising five CSV files. Figure 2 shows the relationships among these tables, and Table 1 summarizes the overall scale of the dataset.

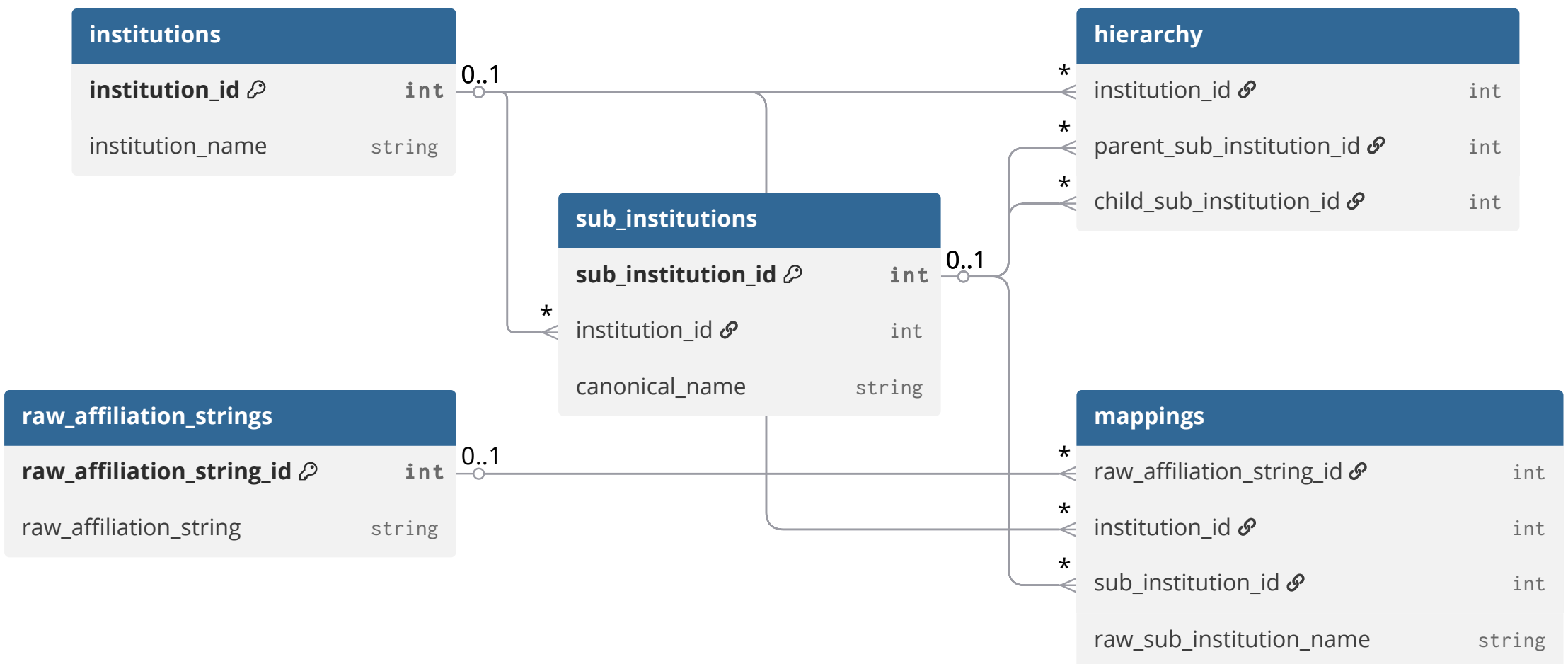

**Fig. 2** Entity-relationship diagram of the OpenSubAffil dataset. Arrows indicate foreign key relationships. The sub_institutions table serves as the central entity linking the mapping and hierarchy tables.

**Table 1.** Overview of the OpenSubAffil dataset.

| Table name | Records | Description |
|---|---|---|
| institutions | 18,635 | Top-level educational institution |
| sub_institutions | 638,843 | Disambiguated sub-institutional entities |
| raw_affiliation_strings | 39,932,792 | Raw affiliation strings |
| mappings | 54,116,686 | Raw affiliation to sub-institution entity mappings |
| hierarchy | 638,843 | Parent-child hierarchical relationships |

The **institutions** table (Table 2) serves as the institution dimension table, containing the 18,635 educational institutions included in the dataset. Each record provides an OpenAlex institution identifier and institution name. All institution identifiers can be linked to ROR identifiers through the OpenAlex API for further enrichment.

**Table 2.** Field description of the institutions table.

| Field | Type | Description |
|---|---|---|
| institution_id | Integer | OpenAlex institution identifier (the prefix "I" is removed), e.g., 37461747. |
| institution_name | String | Name of the top-level institution, e.g., Wuhan University. |

The **sub_institutions** table (Table 3) is the sub-institution dimension table, containing 638,843 disambiguated sub-institutional entities. Each entity is assigned a unique

sub_institution_id, defined over the unique combination of institution_id and canonical_name. This identifier serves as the primary key for linking to both the mapping and hierarchy tables.

**Table 3.** Field description of the sub_institutions table.

| Field | Type | Description |
|---|---|---|
| sub_institution_id | Integer | Unique identifier for the sub-institutional entity, e.g., 76255. |
| institution_id | Integer | OpenAlex institution identifier (the prefix "I" is removed), e.g., 37461747. |
| canonical_name | String | Disambiguated canonical name, e.g., school of information management. |

The **raw_affiliation_strings** table (Table 4) provides the original affiliation text for the 39,932,792 raw affiliation strings involved in the final output. The raw_affiliation_string_id corresponds to the persistent identifier maintained in the CWTS version of OpenAlex, as described in the Methods section.

**Table 4.** Field description of the raw_affiliation_strings table.

| Field | Type | Description |
|---|---|---|
| raw_affiliation_string_id | Integer | Persistent identifier for the raw affiliation string, maintained by CWTS, e.g., 65442162. |
| raw_affiliation_string | String | Original affiliation text as recorded in OpenAlex, e.g., School of Information Management, Wuhan University, Wuhan 430072, China |

The **mappings** table (Table 5) is the core fact table of the dataset, containing 54,116,686 records that link each raw affiliation string to its recognized sub-institutional entities. Each row includes the raw_sub_institution_name, the original sub-institutional name extracted by the NER model, preserving the mapping between raw surface forms and their disambiguated canonical names. A single raw affiliation string may map to multiple sub-institutional entities when the NER model identifies more than one sub-unit within the string.

**Table 5.** Field description of the mappings table.

| Field | Type | Description |
|---|---|---|
| raw_affiliation_string_id | Integer | Raw affiliation string identifier (foreign key). |
| institution_id | Integer | OpenAlex institution identifier (foreign key). |
| sub_institution_id | Integer | Sub-institutional entity identifier (foreign key). |
| raw_sub_institution_name | String | Sub-institutional name as extracted by the NER model, e.g., Sch Inf Mgmt. |

The **hierarchy** table (Table 6) records 638,843 parent-child relationships among sub-institutional entities within each institution. When parent_sub_institution_id is empty, the child entity is a first-level unit directly subordinate to the top-level institution. Non-empty values indicate nested relationships between sub-institutional entities, e.g., a department within a school. Each child has exactly one parent, forming a tree structure.

**Table 6.** Field description of the hierarchy table.

| Field | Type | Description |
|---|---|---|
| institution_id | Integer | OpenAlex institution identifier (foreign key). |
| parent_sub_institution_id | Integer | Parent entity identifier (foreign key). Null indicates a first-level subunit directly subordinate to the top-level institution. |
| child_sub_institution_id | Integer | Child entity identifier (foreign key). |

## Technical Validation

This section presents a systematic validation of the OpenSubAffil dataset. We evaluated the dataset along two dimensions: (i) the quality of sub-institutional name disambiguation and (ii) the accuracy of the reconstructed hierarchical relationships. Since no gold standard exists for sub-institutional entities, we assessed the disambiguation quality through expert annotation on a stratified sample of the data. For hierarchy reconstruction, we benchmarked against Wikidata-derived institutional hierarchies as the primary evaluation, and against the GERiT directory of German research institutions as a supplementary evaluation.

### Sub-institutional name disambiguation quality

***Evaluation design***. We evaluated the disambiguation results by assessing the extent to which a cluster groups variant names of the same sub-institutional unit, and whether any legitimate variants were missed. Essentially, the former measures precision, while the latter measures recall.

***Sampling and annotation strategy***. As no gold standard exists for sub-institutional entities, we rely on expert annotation over a stratified sample. To ensure a representative coverage, we first stratified the 18,635 top-level institutions into three tiers by their number of disambiguated sub-institutional entities: large (top 10%), medium (10% - 50%), and small (bottom 50%), and randomly selected 10 institutions from each tier. Within each selected institution, we further stratified the canonical entities by the number of name variants into three tiers using the same percentile thresholds, and sampled 10 entities per institution. Entities with only one associated name variant were excluded. This yielded 300 canonical entities across 30 institutions for evaluation.

For each sampled entity, we constructed an evaluation set comprising two types of name variants. The first type (original members) comprises sub-institutional names that the clustering algorithm grouped under this canonical entity. The second type (candidates) involves names that exhibit an above-threshold textual similarity to the sampled entity but were not merged by the algorithm. The inclusion of the second-type names enables evaluation of recall, as they represent potential under-merging cases. In total, the evaluation set contains 2,817 original members and 527 candidate members across the 300 sampled entities. We note that the candidate pool is bounded by the textual similarity threshold. Variants that fall below this threshold are not surfaced as candidates and therefore cannot be counted as false negatives. The reported recall should thus be interpreted as recall within the similarity-reachable space and may overstate recall relative to the full population of legitimate variants.

Each name variant was independently evaluated by 2 experts, who judged whether it refers to the same sub-institutional unit as the canonical name (labeled 1) or not (labeled 0). Annotators were instructed to focus on the core part of the sub-institutional names. The two annotators achieved a raw agreement of 90.22% and a Cohen's kappa of 0.67, indicating substantial agreement[40]. To produce final consensus labels, we adopted a conservative strategy. For original members, a name is labeled as incorrectly merged if either annotator assigns a label of 0. For candidate members, a name is considered a missed variant if either annotator assigns a label of 1. This design ensures that both false positives and false negatives are counted generously under the adjudication rule, yielding conservative estimates given the candidate pool. The recall estimate remains bounded above by the candidate-sampling threshold described earlier.

***Evaluation metrics***. Based on the consensus labels, we computed precision, recall, and F1 score. These metrics are defined as:

$$Precision = \frac{TP}{TP + FP} \tag{7}$$

$$Recall = \frac{TP}{TP + FN} \tag{8}$$

$$F1 - Score = 2 \times \frac{Precision \times Recall}{Precision + Recall} \tag{9}$$

A true positive (TP) is an original member confirmed as correctly grouped. A false positive (FP) is an original member judged as incorrectly merged. A false negative (FN) is a candidate member judged as the same as the canonical entity but not merged by the algorithm. We report both micro and macro-averaged metrics.

***Results***. Table 7 reports the overall disambiguation quality. The evaluation yields a micro precision of 0.853, a micro recall of 0.934, and a micro F1 of 0.892. The macro-averaged results are also consistent (precision = 0.884, recall = 0.941, F1 = 0.888). These results indicate that the disambiguation pipeline produces high-quality clusters overall. Notably, the recall consistently exceeds precision across all subgroups, suggesting that the pipeline errs slightly toward over-merging rather than under-merging, which is desirable for downstream applications where missing a valid variant is more costly than including a borderline case.

Disaggregation by institution size tier reveals a precision-recall trade-off related to institutional complexity. Large institutions exhibit the highest micro recall (0.960) but the lowest micro precision (0.807), indicating that institutions with more subunits tend to have more ambiguous and overlapping names. This may increase the risk of incorrect merging while the algorithm still captures most variants. Medium-sized institutions achieve the best balance across all metrics (micro F1 = 0.917, macro F1 = 0.896). A similar pattern is observed when disaggregated by entity name variant tier. Entities with many members achieve the highest micro recall (0.962) but the lowest micro precision (0.838), as larger groups are more likely to include borderline cases. Entities with fewer members show higher micro precision (0.857) but lower micro recall (0.839).

**Table 7.** Disambiguation quality evaluation results.

| **Grouping** | **Subset** | **Entities** | **Micro P** | **Micro R** | **Micro F1** | **Macro P** | **Macro R** | **Macro F1** |
|---|---|---|---|---|---|---|---|---|
| Overall | / | 300 | 0.853 | 0.934 | 0.892 | 0.884 | 0.941 | 0.888 |
| Institution | Large | 100 | 0.807 | 0.960 | 0.877 | 0.855 | 0.962 | 0.885 |

| Grouping | Subset | Entities | Micro P | Micro R | Micro F1 | Macro P | Macro R | Macro F1 |
|---|---|---|---|---|---|---|---|---|
| | Medium | 100 | 0.907 | 0.928 | 0.917 | 0.908 | 0.929 | 0.896 |
| | Small | 100 | 0.850 | 0.878 | 0.864 | 0.889 | 0.932 | 0.882 |
| Entity | Large | 100 | 0.838 | 0.962 | 0.896 | 0.844 | 0.965 | 0.884 |
| | Medium | 100 | 0.892 | 0.919 | 0.905 | 0.903 | 0.944 | 0.901 |
| | Small | 100 | 0.857 | 0.839 | 0.848 | 0.900 | 0.918 | 0.878 |

Among the error cases, the most common false positives involve entities with partially overlapping core names at different granularities (e.g., Department of Biology merged with Department of Integrative Biology), while false negatives typically involve heavily abbreviated or truncated variants that fell below the similarity threshold (e.g., School of Tourism and Environment not merged with Tourism School).

## Sub-institutional hierarchy reconstruction quality

***Evaluation design***. Evaluating the accuracy of reconstructed hierarchies requires annotated datasets of known parent-child relationships among sub-institutional entities. While such information may be available on university websites, collecting it across thousands of institutions is prohibitively labor-intensive and difficult to scale. We therefore constructed a benchmark from Wikidata and additionally used the GERiT directory of German research institutions as a supplementary evaluation.

***Benchmark construction***. We leveraged the fact that Wikidata contains organizational structure information for many academic institutions through its “part of” (P361) property, and that these institutions can be linked to OpenAlex via their OpenAlex identifiers (P10283). To construct the benchmark, we queried the Wikidata SPARQL endpoint for organizational hierarchies of higher education institutions across the 10 most active countries in scientific research, including China, the United States, the United Kingdom, India, Germany, Italy, Japan, Canada, Spain, and Australia. For each country, we identified root nodes as instances of “university” (Q3918) or its subclasses, and extracted all direct parent-child edges where the child is typed as faculty (Q180958), college (Q189004), or academic department (Q2467461). Transitive reduction was applied to remove redundant edges. The parameterized SPARQL query used to retrieve the Wikidata hierarchy records is provided in the validation directory of our code repository. To ensure compatibility with OpenSubAffil, we cleaned the extracted names by removing institution-level prefixes commonly present in Wikidata labels (e.g., University of Oxford Faculty of History was shortened to Faculty of History), and filtered to English-language entries using language detection. After filtering to institutions present in OpenSubAffil, the final Wikidata benchmark contains 6,790 hierarchical edges across 705 institutions.

As a supplementary benchmark, we used the GERiT (German Research Institutions) directory, which contains curated organizational hierarchies maintained by the German Research Foundation (DFG). The version used in this study was obtained directly from the DFG through email correspondence with Dr. Jürgen Güdler and Dr. Martin Schäfer. After applying the same cleaning and matching procedure, the final GERiT benchmark used in this study contains 15,302 edges across 296 institutions. This benchmark enables a direct comparison with Backes et al.[35], one of the few prior works that have attempted to reconstruct the sub-institutional-level hierarchy.

***Evaluation metrics***. We evaluated the hierarchy quality by comparing reconstructed parent-child relationships against the benchmarks. Since the names of the institutions and subunits may differ between the benchmarks and OpenSubAffil, we used fuzzy name matching (token set ratio ≥ 85) to align nodes across sources. For each edge in the benchmarks, we independently matched its parent name and child name to our node pool using fuzzy matching. If both names matched successfully, we then checked whether the matched node pair exists as an edge in our predictions. If so, it was counted as a true positive for recall. Recall is defined as:

$$Recall = \frac{matched\ edges\ in\ OpenSubAffil}{total\ edges\ in\ the\ benchmark} \quad (10)$$

For precision, we performed the reverse: for each reconstructed edge, we independently matched its parent and child names to the benchmark. If both names matched successfully, we then checked whether the matched node pair exists as an edge in the gold standard. However, not all reconstructed edges can be evaluated in this way. If either endpoint has no corresponding node in the benchmark, the correctness of the reconstructed edge cannot be determined. Therefore, we restrict the precision calculation to reconstructed edges where both nodes can be matched, which we term comparable edges. Precision is defined as:

$$Precision = \frac{matched\ edges\ in\ the\ benchmark}{comparable\ edges\ in\ OpenSubAffil} \quad (11)$$

We evaluated the quality of sub-institutional hierarchy at two levels. The all-edges evaluation includes every parent-child relationship. The nested evaluation excludes edges from the top-level institutions and retains only relationships between two sub-institutional entities (e.g., faculty to department). This kind of relationship is more difficult to reconstruct from raw affiliation strings. We report both micro and macro-averaged (across institutions) metrics in the next section.

***Results***. Table 8 reports the hierarchy evaluation results against the Wikidata benchmark. The all-edges evaluation shows a micro precision of 0.830 and a micro recall of 0.777 (F1 = 0.803). The macro-averaged metrics are notably higher, with precision 0.940 and recall 0.793, suggesting that most institutions achieve high accuracy within the benchmark's scope and that the lower micro scores are driven by a small number of institutions with complex structures. Disaggregation by institution size confirms this pattern. Smaller institutions achieve substantially higher precision compared to larger institutions, while recall remains relatively stable across all tiers.

The nested evaluation, which retains only relationships between two sub-institutional entities across the 60 institutions that contain such edges in the Wikidata benchmark, yields a micro precision of 0.783 but a substantially lower micro recall of 0.368 (F1 = 0.500). This is expected, as intermediate organizational layers, such as faculties, are frequently omitted in affiliation strings, leaving limited co-occurrence evidence for the scoring function. Nonetheless, the high precision at this level indicates that when the algorithm does predict a nested relationship, it is usually correct.

**Table 8.** Sub-institutional hierarchy evaluation results against the Wikidata benchmark.

| Level | Subset | Institutions | Micro P | Micro R | Micro F1 | Macro P | Macro R | Macro F1 |
|---|---|---|---|---|---|---|---|---|
| All edges | Overall | 705 | 0.830 | 0.777 | 0.803 | 0.940 | 0.793 | 0.818 |
| | Large | 71 | 0.753 | 0.747 | 0.750 | 0.791 | 0.767 | 0.760 |
| | Medium | 282 | 0.874 | 0.804 | 0.837 | 0.920 | 0.790 | 0.830 |
| | Small | 352 | 0.973 | 0.803 | 0.880 | 0.986 | 0.799 | 0.821 |
| Nested | Overall | 60 | 0.783 | 0.368 | 0.500 | 0.855 | 0.234 | 0.266 |

***Supplementary evaluation against GERiT***. The Wikidata benchmark, while global in scope, is limited in depth: only 60 of the 705 institutions contain nested edges. To provide additional evidence for deeper hierarchies and to enable comparison with prior work[35], we evaluated against GERiT. As shown in Table 9, OpenSubAffil achieves a micro precision of 0.329 and a micro recall of 0.117 at the all-edges level. While these absolute values are substantially lower than those on the Wikidata benchmark, this gap is largely explained by the nature of the GERiT hierarchy itself. GERiT records organizational structures up to seven levels deep, including fine-grained units such as individual chairs and research groups. Authors affiliated with these units, however, rarely include all intermediate levels in their affiliation strings. For instance (Figure 3), GERiT records the path "Trier University → Department IV – Economics, Social Science, Computer Science and Mathematics → Subject Unit: Economics → Chair of Empirical Economics Research". In practice, however, authors typically write simply "Department of Economics, Trier University". This not only omits multiple intermediate levels but also employs a naming convention that is entirely absent from the GERiT hierarchy. Such intermediate layers are therefore difficult to recover for methods that rely on co-occurrence evidence in affiliation strings, including ours.

These challenges are consistent with the findings of Backes et al.[35], who introduced the task of unsupervised hierarchical affiliation resolution and evaluated their method on three manually annotated German universities (including Trier). Using a pairwise dominance protocol that counts all ancestor-descendant relations, they reported 0-4% recall and 0-30% precision. Our method, evaluated across 296 institutions using direct parent-child edges as the evaluation unit, achieves 11.7% micro recall and 32.9% micro precision.

**Table 9.** Sub-institutional hierarchy evaluation results against the GERiT benchmark.

| Method | Level | Inst. | Micro P | Micro R | Micro F1 | Macro P | Macro R | Macro F1 |
|---|---|---|---|---|---|---|---|---|
| OpenSubAffil | All edges | 296 | 0.329 | 0.117 | 0.173 | 0.610 | 0.237 | 0.304 |
| OpenSubAffil | Nested | 296 | 0.387 | 0.048 | 0.085 | 0.498 | 0.027 | 0.045 |
| Backes et al. [35] | All edges | 3 | 0-0.30 | 0-0.04 | - | - | - | - |

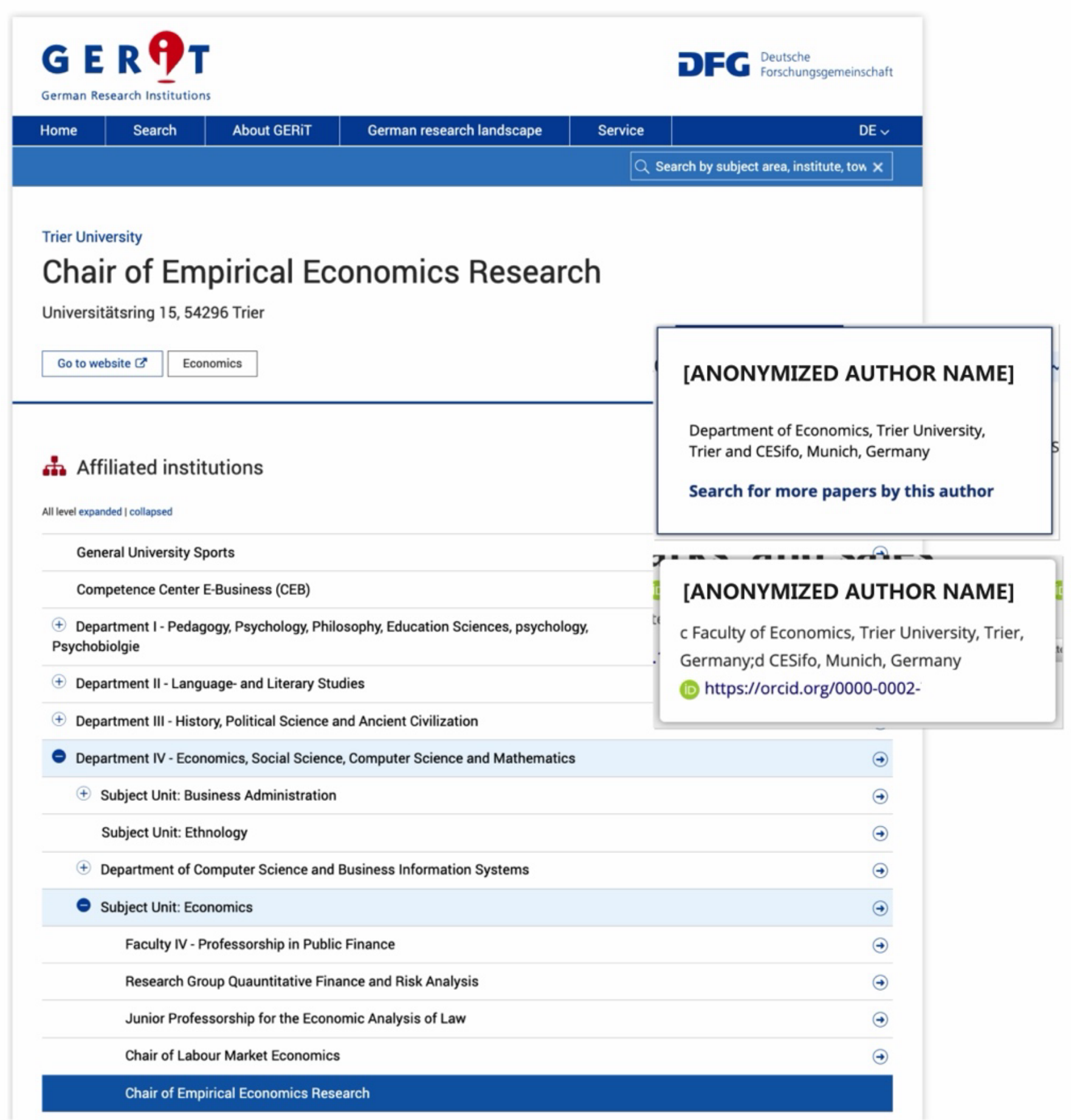

**Fig. 3** Illustration of the gap between organizational hierarchies recorded in GERiT and the affiliation strings written by authors.

Overall, the validation results show that OpenSubAffil provides reliable sub-institutional data across two dimensions. The disambiguation pipeline achieves a macro F1 of 0.888, with high recall suggesting that most legitimate name variants are successfully captured. The reconstructed hierarchies align well with the Wikidata benchmark at the all-edges level (macro F1 = 0.818) and exhibit high precision, indicating that the reconstructed relationships capture meaningful hierarchical structure, although nested relationships remain more difficult to recover reliably.

## Usage Notes

OpenSubAffil is designed to support research and applications that require sub-institutional granularity, such as departmental-level research assessment, within-institution collaboration analysis, fine-grained tracking of knowledge flows across organizational units, and institutional management and policymaking.

The dataset is publicly available at https://doi.org/10.5281/zenodo.19602782. It consists of five CSV files that can be processed with standard data analysis tools such as Python or R. The raw_affiliation_string_id in the dataset corresponds to the persistent identifier maintained in the CWTS version of OpenAlex, which is available on Google BigQuery. This identifier is not present in the official OpenAlex JSON data dump. Users wishing to link OpenSubAffil records back to the original affiliation strings and their associated publications should use the CWTS version. All institution identifiers correspond to OpenAlex institution IDs and can be linked to ROR identifiers through the OpenAlex API for further enrichment.

Several considerations should be noted when using the dataset. First, OpenSubAffil covers only educational institutions. Sub-institutional structures for other institution types are not included, as the pipeline's design assumptions are tailored to educational organizations.

Second, the dataset is derived from English-language affiliation strings. Institutions in non-English-speaking countries are included if their authors publish with English-language affiliations, but coverage of sub-institutional entities may be less complete for these institutions.

Third, the hierarchical relationships are reconstructed from co-occurrence patterns in affiliation strings. Intermediate organizational layers that are rarely mentioned in affiliations may be underrepresented. Users requiring complete organizational trees for specific institutions may need to supplement the dataset with information from institutional websites or national directories.

Fourth, the disambiguation is performed independently within each top-level institution. The same canonical name (e.g., "department of computer science") may appear under different institutions and refers to distinct entities in each case. The sub_institution_id field should be used for cross-institutional analyses to avoid conflation.

Fifth, the canonical_name values are normalized to lowercase during disambiguation and should be treated as normalized labels rather than display-ready names. Consequently, the original capitalization of acronyms and Roman numerals is not preserved. Users requiring the original casing can refer to the raw_sub_institution_name field in the mappings table, which retains the form extracted from the affiliation string.

## Data Availability

The OpenSubAffil dataset is publicly available at https://doi.org/10.5281/zenodo.19602782 under a CC BY 4.0 license. The two expert-annotation files, the processed Wikidata benchmark, and the OpenAlex education-institution ROR mapping used for validation are available in the accompanying code repository. The GERiT source data are not redistributed because they were provided by the DFG under a data-use condition limiting their use to the specified research purpose.

## Code Availability

The source code is available at https://github.com/tomleung1996/OpenSubAffil.

## Funding

This work was funded by the National Natural Science Foundation of China (Grant Nos. 72504210, L2524097, 72474159) and China Postdoctoral Science Foundation (Grant No. 2025M773208).

## Acknowledgements

The authors would like to express their sincere gratitude to Dr. Jürgen Güdler and Dr. Martin Schäfer at the Deutsche Forschungsgemeinschaft (DFG) for generously providing the GERiT data, and to Dr. Nees Jan van Eck at the Centre for Science and Technology Studies (CWTS), Leiden University, for providing the CWTS relational version of the OpenAlex database and for his valuable feedback on this work. The first author is happy to celebrate his birthday with this publication.